# Digital Transformation of Nature Tourism

Raul Enrique Rodriguez Luna - José Luis Rosenstiehl Martinez

November 8, 2022

**Abstract**

The purpose of this article is to explore the digital behavior of nature tourism SMEs in the department of Magdalena-Colombia, hereinafter referred to as the region. In this sense, the concept of endogenization as an evolutionary mechanism refers to the application of the discrete choice model as an engine of analysis for the variables to be studied within the model. The type of study was quantitative, of correlational level, the sample was 386 agents of the tourism chain; a survey-type instrument with five factors and a Likert-type scale was used for data collection. For the extraction of the factors, a confirmatory factor analysis was used, using structural equations, followed by a discrete choice model and then the analysis of the results. Among the main findings are that the SMEs in the tourism chain that tried to incorporate Big Data activities in the decision-making processes, have greater chances of success in the digital transformation, in addition, statistical evidence was found that the training of staff in Data Science, contributes significantly to the marketing and commercialization processes within the SME in this region.

## 1 Introduction

Digital transformation refers to an abstract model that attempts to analyze some modern factors of the so-called digital transformation in an economic activity that has been restructuring the types of work and the form of service provision in the different economic sector.

In this regard, Bhiman Willcock (2014) argue that the costs of operational and administrative processes prevent or restrict the digital transformation for SMEs, although it is true that tourism chains are information intensive, many of these SMEs cannot handle large volumes of data (p. 19).

However, McAfee Brynjolfsson (2012) have shown that the variables, volume, speed and variety of data are the main problems faced by companies for such digital transformation and consequently the low rates in the generation of value. On the other hand, it is expected that the exponential increase in data will overflow the capacity of SMEs' management information systems and consequently the staff will not be able to make decisions in the face of the challenges of demand, as in Desjardins, (2019) who estimated that, by 2025, 463 exabytes of data will be created every day globally.

In this sense, the thesis that this article supports are, first, the probability of success of a digital transformation in nature tourism SMEs in the department of Magdalena would be associated with five factors, in their order, first, the incorporation of Big Data in decision making, second the training of personnel in Big Data tools, third, outsourcing processes in Big Data, fourth the demand and fifth the strength of e-commerce.In this regard, (Grover et al., 2018)state that, for a company to have positive returns, it is essential to justify the investment in Big Data tools to extract, process and understand the data in the decision-making process.Furthermore, Müller, O., Fay, M., Brocke, J., (2018) argue that "to optimize the use of available resources in the value chain and try to control endogenous variables within the organiza- tional structure, it is necessary to intervene in human capital from training" (p. 6). On the other hand, the digital transformation must be seen from the organizational performance and find arguments in favor of endogenization for the achievement of such transformation (Gunasekaran et al., 2017). That is why, with this article we seek to analyze some factors that could give lights to endogenize digital transformation, finally the research question focuses on how to endogenize digital transformation in nature tourism SMEs within.Endogenization as a mechanism value chains in the region? for this, some variables will be modeled from prediction models with the support of Big Data theory, value chains and human capital theories, on the other hand this paper contributes with the theory of value chains, following the common thread for purposes of analysis of the specialized and non-specialized reader; the work consists of seven parts starting with this introduction followed by theoretical references, working hypotheses, methodology, results, discussion finally concludes.

## 2  Literature Review

Digital transformation is the concept that has gained momentum in recent years, not only because it seeks to measure the use of ICTs in organizations, but also because it is an essential element in everyday life, affecting all dimensions that involve individuals and companies (Rodriguez and Bibriesca 2019, p. 4). In this sense, Morakanyane et al. (2017) describe digital transformation as an evolutionary process that leverages digital capabilities and technologies to enable business models, operational processes and consumer experiences that generate value. Digital transformation strategies are innovation strategies, which focus on the transformation of products, processes and other organizational aspects, thanks to new technologies. This includes user interaction with technology as an integral part of the product or service and allows defining new business models together (Matt et al., 2015). In this way, digital transformation is the integration of technologies to generate value for customers, according to (Hofmann, 2017).

In this way,Varian (2013);Chen,2012);(Chen, C., Zhang, C. 2014).In this regard Varian (2013)states that,Big Data are tools to analyze data, which can be loaded from statistical packages, process information and proceed with data interpretation and initiate decision processes in the organization(p.4).

On the other hand, (Chen, 2012);(Chen, C., Zhang, C. 2014) Big Data are complex data sets that classical tools from the sciences or coming from classical statistics for data processing are not optimal to handle them.

This is why Big Data applications make it possible to increase the performance and productivity of tourism chains and their insertion in the international marke.

In this regard,Morakanyane et al. (2017); in this work this phenomenon is relativized from outsourcing processes, value towards the tourism chain, data-driven innovations, sales in the marketing process, demand, health, performance measures, failures in managerial decisions due to lack of predictive models, productivity, forecasting to minimize errors in decision-making processes, e-commerce services and value chains. However, on the subject of outsourcing processes, these enable digital transformation and generate value to the chain, however, "the use of Big Data is not clear what types of organizations are benefiting from innovation associated with Big Data and open data" (Chiang et al., 2011, pp.219-234); in this context it is expected that companies will benefit the most from these data-driven innovations. Likewise, sales in the marketing process, "demand plays a key aspect in management, therefore, it is important demand planning in terms of investment in technologies for appropriation" (Addo, 2016, pp. 528-543).Although, Big Data is increasingly becoming a business force, to discuss problems related to analysis, trends, this could regenerate negative consequences on health (Alfall et al., 2014) and (Chen and Zhang, 2014, pp. 314-347). In sum, "the relationship between employee and chain integration allow for improved performance measures", (Hazen, 2014, pp. 72-80). On the other hand, failures in managerial decisions often occur due to lack of predictive models in the chain, which often leads to errors and productivity losses given the large volumes of data and the speed with which this occurs, in this sense, Hazen (2014) argues that Big Data makes it possible to define predictive models that guarantee an optimal forecast to minimize errors in decision-making processes, consequently, it is necessary to monitor structured or unstructured data, for socioeconomic, technical, or financial purposes in an increasingly dynamic and competitive context. Therefore, in each and every one of the actions that are undertaken in function of the achievement of the chain, the function of Big Data in the digital transformation plays a fundamental role in the disposition and action, in the understanding that the true objectives of the chain, are only achieved if human capital trained in high performance productive operations is at hand, such as outsourcing processes, optimal sales performance measured by investment in Big Data and optimal e-commerce services and predictive models to prevent failures by incorporating Big Data techniques as indicated by the studies of (Chiang et al., 2011).

However, on the subject of outsourcing processes, these enable digital transformation and generate value to the chain, however, "the use of Big Data is not clear what types of organizations are benefiting from innovation associated with Big Data and open data" (Chiang et al., 2011, pp.219-234); in this context it is expected that companies will benefit the most from these data-driven innovations. Likewise, sales in the marketing process,"demand plays a key aspect in management, therefore, it is important demand planning in terms of investment in technologies for appropriation" (Addo, 2016, pp. 528-543).Although, Big Data is increasingly becoming a business force, to discuss problems related to analysis, trends, this could regeneratenegative consequences on health (Alfall et al., 2014) and (Chen and Zhang, 2014, pp. 314-347). In sum, "the relationship between employee and chain integration allow for improved performance measures", (Hazen, 2014, pp. 72-80). On the other hand, failures in managerial decisions often occur due to lack of predictive models in the chain, which often leads to errors and productivity losses given the large volumes of data and the speed with which this occurs, in this sense, Hazen (2014) argues that Big Data makes it possible to define predictive models that guarantee an optimal forecast to minimize errors in decision-making processes, consequently, it is necessary to monitor structured or unstructured data, for socioeconomic, technical, or financial purposes in an increasingly dynamic and competitive context. Therefore, in each and every one of the actions that are undertaken in function of the achievement of the chain, the function of Big Data in the digital transformation plays a fundamental role in the disposition and action, in the understanding that the true objectives of the chain, are only achieved if human capital trained in high performance productive operations is at hand, such as outsourcing processes, optimal sales performance measured by investment in Big Data and optimal e-commerce services and predictive models to prevent failures by incorporating Big Data techniques as indicated by the studies of (Chiang et al. , 2011, pp.219-234 ); Dobre Xhafa (2014); (Hann Steurer, 1996) and(Jagielska Jaworski, 1996, pp. 77-82). On the other hand, the Big Data brings several advantages for companies, among others, transparency of information for the chain, shared vision and improvement in decisions, which guarantees their customers better levels in service delivery, however, Wu et al (2019) express that, the analysis of records based on Big Data process data of large volumes and speeds and consequently, it is difficult for human talent to predict all economic orother phenomena, in this regard, Zhang, G., Berardi, V. (2001) state that "neural networks improve time series prediction performance and enable forecasting methods to improve the decision making process" (pp. 652-664). In the same way, Pietro (2002) suggests that the use of predictive models is not recommended if there are workers with low human capital in mathematical modeling, in other words, low levels of human capital are correlated with high levels of errors in decision making. In this way, prediction techniques improve the price forecasts of the products in the chain when large volumes of data are used, in this order Ortiz, Cabrera López (2013) explain that, the use of Garch models and neural networks as an alternative method in the calculation of future price forecasts, however the use of neural networks as a reliable alternative method in the analysis of time series significantly improves predictions minimizing failures or errors in the decision making process, consequently, the use of Big Data contributes with a competitive advantage of the companies that belong to the supply chain as suggested by (Bhimani Willcocks 2014 pp. 469-490 (2020); Waller and Fawcett (2013); Zhang and Berardi (2001) and Zhong et al. (2016) are sufficient and necessary conditions for the achievement of a sustainable digital transformation.

## 3 Methodology

The work is quantitative, descriptive and explanatory, with a correlational scope, since explanatory research enables cause-effect relationships, based on the use of hypothesis testing. The field design allowed the collection and description of the data, these come from managers of high and middle management in the tourism chains of the department, for this, the work is supported by the structured survey which is framed in the digital transformation taken as a moderating variable the use of Big Data in decision making as a proxy variable of the digital transformation of nature tourism SMEs and consequently of the tourism chain for the region. The theory developed by Big Data was used as a proxy of the digital transformation, four factors were taken as explanatory variables from the confirmatory factor analysis, a process not shown in this article, these factors are, personnel training, personnel outsourcing processes in the chain, sales per service and prediction techniques; the sample consisted of 386 agents of the tourism chain in Magdalena, then the final subjects of the different strata were randomly selected proportionally, under a confidence level of 95

$F(s) = 1/ (1 + e^{-s})$, $-¡ s ¡$ .

In fact, the dependent variable $F(s)$ use of Big Data in the decision-making process in the commercial link of SMEs in the chain was used as a proxy variable for digital transformation, following the work of Waller Fawcett (2013), which supports a positive association between the availability of Big Data and its use in decision-making in chain management and consequently increasing the performance of these chains. Economic theory predicts that the training of personnel is significant for the achievement of digital transformation; in this respect, to the extent that personnel are incorporated into business processes, they improve predictions, which makes the process sustainable (Grover et al., 2018, pp. 388-423).On the other hand, in terms of personnel outsourcing processes, the hypothesis put forward suggests that Big Data allows improving operational efficiency and generating value for the chain, following the work of Wilkin et al, (As for the sales factor in the marketing process, the thesis here suggests that, although demand plays a key aspect in the management of the chain, its planning is more relevant if it is shown in terms of investment in technologies for appropriation of Big Data (Varian, 2013 pp. 3-27). Finally, the last variable incorporated in the model is related to forecasting techniques for demand analysis, based on forecasting models if large volumes of data are used (Hann and Steurer, 1996).

# 4  Conclusions

To appropriate the results of the model and removegenerality to the data the sample takes the form y = (y1, y2, ..., yn) which consisted of a vector of dimension n $\times$ 1 and a vector of dimension k $\times$ 1, xi = (xi1, ..., xik), of covariates or explanatory factors for each of the i = 1, ..., n, with logistic regression we estimated the probability that the decision processes succeed from each of the i = 1, ..., ..., n. xik), of covariates or explanatory factors for each of the i = 1, .........., n, with the logistic regression we estimated the probability that the decision processes will be successful from the use and utilization of Big Data and consequently there is the possibility of a digital transformation.

In relation to the latter,Wilkin et al, (2020) support a positive association between the availability of Big Data and its use in decision making in tourism chain management and consequently increasing the performance of these. On the other hand, value chains have a positive association with digital transformation from the customer's point of view, since these are theoretical models that attempt to describe the iterations between the activities of a business organization to generate value for the end customer, and for Anguiano and Pancorbo (2008), these are a set of interrelated activities whose main objective is to obtain a competitive advantage. Finally, Landon and Michael (2016, pp. 92-121) "the best and simplest way to approach data science is probably to look at what data scientists do." This might suggest two issues: first, higher education in Latin America is in debt and has only recently begun to introduce data science programs, which means that higher education must adjust quickly to these market needs in terms of training, and second, it is necessary to adjust the physical capital of SMEs in these chains in some way, possibly with access to banking through loans, and to ensure by some mechanism that this investment is directed towards the technical progress of this type of organization.

The resulting model took the functional form
F(s)= 1/ (1 + e -s),(- ¡ s ¡ ) for which the symmetric logistic distribution was used,that is F (k - s)= F(k + s),for a constant k and for any s;=(1, k) is a vector of dimension k $\times$ 1 formed by the parameters to be estimated, and which represents the effect of each of the covariates in the model.

Subsequently, the regression coefficients and other parameters were obtained, using Eviews and SPSS v.21, the resulting model.The adjusted model was constructed from the survey-type instrument and the confirmatory factor analysis to explain the dependent variable (E) which were found to be statistically significant.

According to the analysis carried out, it was found that digital transformation as an endogenous variable needed to be studied based on the use and availability of Big Data tools in the chain, dichotomized as one (1) if the tourism link has tools for handling large volumes of data and zero if not, it was found that only 2

Endogenization as an evolutionary mechanism for the digital transformation of nature tourism Although the results, are not as conclusive as one would like, this could indicate that, there is very little likelihood of using Big Data in the tourism chain in this region and consequently deficiencies and low levels of digital transformation. Again, this is confirmed by the R-squared of the model (0.34). In this regard, Pietro (2002) suggests that the use of Big Data as a support for digital transformation and decision-making processes reduces costs and improves efficiency; however, there should be no illusions about this policy. On the other hand, although the volume of data will always exist in all economies and many times it is not known what to do with it, it constitutes a type of negative externality for digital transformation, but the important thing according to Chianget al. (2011) is to try to reduce the volume of data to the level where the externalities produced by this sector proportionally compensate the economic damage it generates to the formal industry. In general terms, the training of personnel in Big Data tools Independent variable (x) was studied based on the type of training of chain personnel with training in science and engineering of middle and upper management; it was found that 10 percent of managers at this level have specialized training in Data Science which allows complex problems in the tourism chain to be solved. Another important result is that outsourcing in Big Data as an independent variable (y) was studied based on the number of external people hired for information management activities with large volumes of data that must be processed at high speeds to generate value and profitability in the process of business decisions; statistical evidence of reduction of human resources in the chain was found to be around 1 percent, for the sample studied, suggesting few outsourcing processes in this area. However, in terms of demand, nose evidenced increases in sales in percentage terms from the use of Big Data tools in the process of marketing and distribution of goods and services.

Regarding the subject of e-commerce, statistical evidence was found that suggests that, companies in the chain that incorporate e-commerce in the provision of tourism services improved the decision processes in the marketing of their products as was reasonable, in those companies in the chains where e-commerce was incorporated, it allowed a boost to the process, following the contributions of Alfall et al. (2014). These suggest that, delivery generates positive effects to increase the use and integration of the supply chain, help to achieve supplier and customer integration; as a result, companies should strive to achieve employee commitment to optimize e-commerce and internal integration, as it enables the use of Big Data tools in the decision-making process. In conclusion, the findings show a positive association between the variables personnel training, personnel outsourcing, e-commerce and the use of Big Data models and the performance of the chain studied; there is also a mediating effect on the performance of the chain and consequently on the digital transformation of SMEs of nature in the region, as Desjardins (2019) stated that data science, predictive analysis and Big Data are a revolution that transforms the design and management of value chains.

# 5 Implications for theory and practice

Statistical evidence was found in favor of the hypothesis put forward in this work, regarding the training of personnel it is important to train them in tools for the use of Big Data to optimize decision-making processes, so as to intervene in the efficiency of management.For a successful digital transformation it is necessary to incorporate Big Data to optimize commercial management, as well as the incorporation of different types of non-conventional payments, so as to ensure optimal service delivery, so that the chain does not fall behind in terms of technology.

In terms of personnel outsourcing processes, personnel with high training in Big Data and e-commerce are required, so that the chain can face challenges of speed and delivery time for the service. On the other hand, with personnel trained in tools for the use of Big Data and in Data Science for the chain, this contributes to the use of Big Data for decision making, since it allows complex problems to be solved, to go beyond the sector's predictions. Another important conclusion is that, when we inquired with managers, we learned that the percentage of companies that plan commercial management in the supply chain do not apply outsourcing processes for handling large volumes of data or theoretical prediction models, nor data mining, much less analytics, but have adopted traditional forms for planning the chain, especially in terms of processes with low added value and therefore unsophisticated products, to meet the varying needs of demand. It cannot be said that investment in Big Data will be made in the medium or long term; however, the importance of Big Data for the digital transformation of the chain in the region is clear; on the other hand, the managers surveyed stated that if there were more staff training or subcontracting in the use of Big Data tools, there would be more possibilities of increasing the productivity and competitiveness of SMEs in the region's nature industry. In sum, a large percentage of nature SMEs in the region do not engage in e-commerce, have low levels of technological incorporation in their processes, which indicates that the probability of success is only 21

R-squared from 0.332 to 0.429 these results agree with those ofWilkin et al.,(2020) in finding a significant and positive association between the use of Big Data in chain decision making and chain decision performance. Therefore, greater use of Big Data in chain decision making has a beneficial effect on chain performance. The research has empirically reviewed some of the variables that affect the digital transformation seen from the use of Big Data in decision-making processes.In summary, the region's lag in terms of these tools and especially in Latin America, in addition to the above, could be related to the lack of empirical studies reviewed that are only limited to Delphi type studies as in(Zhong et al., 2016, p. 572-591).While understanding our study, this offers a new avenue for future related research it is recognized that, limitations arise from the initial application of the model with the specific use of Big Data for digital transformation in different contexts and provide real-time information that complements and extends current management practices of organizations especially the handling of large volumes of data. Thus, this work contributes to the Big Data theory and the competitive dynamics of the SME nature for a successful digital transformation, and therefore, this work could contribute to future research on the topic of digital transformation from the use of Big Data.

References

(1) Addo, R. Helo, P. (2016). Big data applications in operations/supply-chain management: A literature review. Computers and Industrial Engineering, 101, 528–543. https://doi.org/10.1016/j.cie.2016.09.023

(2) Alfall, L.,Mar´ın, J., Garc´ıa,C. y Medina, L. (2014). An analysis of the direct and mediated effects of employee commitment and supply chain integration on organizational performance. Revista Internacional de Economía de la Producción, 162, 242–257.

(3) Ahiaga, A. (2014). Dealing with construction cost overruns using data mining. Construction Management and Economics,32,682-694. https://doi.org/10.1080/01446193.2014.933854.

(4) Anguiano, A. y Pancorbo, S. (2008). El marketing urbano como herramienta de apoyo a la gestión del turismo urbano de ciudad: estudio de un caso, el patrimonio industrial. Revista ACE,(9), 739-748.

(5) Angus, D. Rintel, J. (2013). Making sense of big text: a visual-first approach for analysing text data using
Leximancer and Discursis International.
Journal of Social Research Methodology, 16, 261-267.

(6) Artigues, A., Cucchietti, F., Tripiana Montes, C., Vicente, D., Calmet, H., Marin, G., Houzeaux, G. Vazquez, M. (2015). Scientific Big Data Visualization: a coupled tools approach.Supercomputing Frontiers and Innovations, 3,4-18. doi: http://dx.doi.org/10.14529/jsfi140301

(7) Arunachalam, D., Kumar, N. Kawalek, J. (2018). Understanding big data analytics capabilities in supply chainmanagement:
unravelling the issues, challenges and implicationsfor practice. Transportation Research Part E: Logistics and Transportation Review, 114, 416–436.

(8) Bhimani, A. Willcocks, L.(2014).Digitization, 'Big Data' and the transformation of accounting information.Accounting and Business Research, 44, 469-490
https://doi.org/10.1080/00014788.2014.910051

(9) Brinch, M. (2018). Understanding the value of big data in supply chain management and its business processes. International Journal of Operations Production Management,38(7), 1589–1614.

(10) Chen, H., Chiang, R. Storey, V. (2012). Business Intelligence and Analytics: From Big Data to Big Impact.MIS Quarterly,36 (4), 1165-1188. doi:10.2307/41703503.

(11) Chen, C. Zhang, C. (2014). Data-intensive applications, challenges, techniques, and technologies: A survey on Big Data. Information Sciences, 275, 314–347. doi: 10.1016/j.ins.2014.01.015.

(12) Chiang, D., Lin, C. Chen, M. (2011). The adaptive approach for storage assignment by mining data of warehouse management system for distribution centers.Enterprise Information Systems, 5,219-234.

(13) Desjardins, J. (2019). How much data is generated each day?News.
https://enewswithoutborders.com/2019/04/23/how-much-data-is-generated-each-day/

(14) Dobre, C. Xhafa, F. (2014). Intelligent services for Big Data science.Future Gener. Comput. Syst, 37, 267-281.

(15) Gunasekaran, A., Papadopoulos, T., Dubey, R., Wamba, S. F., Childe, S. J., Hazen, B. Akter, S. (2017). Big data and predictive analytics for supply chain and organizational performance. Journal of Business Research, 70, 308–317. https://doi.org/10.1016/j.jbusres.2016.08.004

(16) Grover, V., Chiang, R. H., Liang, T. Zhang, D. (2018). Creating strategic business value from big data analytics: a research framework. Journal of Management Information Systems, 35(2), 388–423.
https://doi.org/10.1080/07421222.2018.1451951(17)Hann, T. H. Steurer, E. (1996). Much ado about nothing? Exchange rate forecasting: Neural networks vs. linear models using monthly and weekly data. Neurocomputing, 10, 323–339. https://doi.org/10.1016/0925-2312(95)00137-9

(18) Hazen, L. (2014). Data quality for data science, predictive analytic, and big data in supply chain management: An introduction to the problem and suggestions for research and applications. international Journal of Production Economics, 154, 72-80.

(19) Hofmnn, E. (2017). Big data and supply chain decisions: the impact of volume, variety, and velocity properties on the bullwhip effect. Intenational Journal of Production Research, 55(17), 5108–5126.

(20) Jagielska, I. Jaworski, J. (1996). Neural network for predicting the performance of credit card accounts.Comput Econ, 9,77–82. https://doi.org/10.1007/BF00115693 Endogenización comomecanismo evolutivo para la transformación digital de las Pymes de turismo de naturalezaRaúl Enrique Rodriguez Luna; José Luis Rosenstiehl Martinez137

(21) Kubina, M., Varmus, M. Kubinova, I. (2015). Use of big data for competitive advantage of company. Procedia Economics and Finance,26, 561–565.

(22) Landon, M. Michael, A. (2016). Big Data and Intelligence: Applications, Human Capital, and Education.Journal of Strategic Security,(9),92–121.

(23) Matt, C., Hess, T. Benlian, A. (2015). Digital Transformation Strategies. Business Information Systems Engineering, 57(5), 339–343. DOI 10.1007/s12599-015-0401-5

(24) McAfee, A. Brynjolfsson, E. (2012). Big data: the management revolution.Harvard business review,90,60–128.

(25) Morakanyane, R., Grace, A. O'Reilly, P. (2017). Conceptualizing Digital Transformation in Business Organizations: A Systematic Review of Literature. Digital Transformation –From Connecting Things to Transforming Our Lives. Bled, Slovenia DOI:10.18690/978-961-286-043-1.30

(26) Müller, O., Fay, M. Brocke, J. (2018). The Effect of Big Data and Analytics on Firm Performance: An Econometric Analysis Considering Industry Characteristics. Journal of Management Information Systems, 35, 488-509. https://doi.org/10.1080/07421222.2018.1451955

(27) Nitzl, C., Roldán, J. Cepeda, G. (2016). Mediation Analysis in Partial Least Squares Path Modeling: Helping Researchers Discuss More Sophisticated Models. Econometrics, 116,1849–1864.

(28) Ortiz, F., Cabrera, A. López, F. (2013). Pronóstico de los índices accionarios DAX y SP 500 con redes neuronales diferenciales.Contaduría y Administración, 58, 203-225.

(29) Pietro, G. (2002). Technological change, labor markets, and 'low-skill, low-technology tras'. Technological Forecasting and Social Change, 69,885-895.

(30) Rodriguez, G. Bibriesca, G. (2019). Modelo de Transformación Digital en las empresas. XXXII Congreso Nacional y XVIII Congreso Internacional de Informática y Computación de la ANIEI, Puebla, México.

(31) Varian, H. (2013). Big Data: New Tricks for Econometrics.The Journal of Economic Perspectives, 28,3–27.

(32) Waller, M. Fawcett, S. (2013). Data science, predictive analytics, and big data: a revolution that will transform supply chain design and management.Empirical Supply Topic, 34,77–84.

(33) Wilkin, C., Ferreira, A., Rotaru, K. Gaerlan, L.R. (2020). Big dataprioritization in SCM decision-making:Its role and performance implications.

International Journal of Accounting Information System, 38,pp 100470.

(34) Wu, P., Lu, Z., Zhou, Q., Lei, Z., Li, X., Qiu, M. Hung, P. (2019). Bigdata logs analysis based on seq2seq networks for cognitive Internet of Things. Future Generation Computer Systems, 90,477-488.

(35) Zhang, G. Berardi, V. (2001). Time series forecastingwith neural network ensembles:

An application for exchange rate prediction. Journal of the Operational Society,52,652-664,DOI:10.1057/palgrave.jors.2601

(36)Zhong, R., Newman, S., Huang, G. Lan, S. (2016). Big data for supply chain management in the service and manufacturing sectors: challenges, opportunities, and future perspectives. Computers and Industrial Engineering,101,572 –591.